\documentclass{jaa}
\usepackage{natbib}
\usepackage{graphicx}

\usepackage{hyperref} 
\hypersetup{colorlinks,citecolor=blue,linkcolor=blue,urlcolor=blue}

\newcommand{\arcsec}{\ensuremath{^{\prime\prime}}}
\newcommand{\arcmin}{\ensuremath{^\prime}}
\newcommand{\degr}{\ensuremath{\hbox{$^\circ$ }}}
\usepackage{soul}

\newcommand\chandra{{\it Chandra }}

\newcommand\kms{\ifmmode {\rm~km\ s}^{-1} \else ~km s$^{-1}$\fi}
\newcommand\Hunit{\ifmmode {\rm~km\ s}^{-1}\ {\rm Mpc}^{-1}
        \else ~km s$^{-1}$ Mpc$^{-1}$\fi}
\newcommand\ctssec{\ifmmode {\rm~count\ s}^{-1} \else ~count s$^{-1}$\fi}
\newcommand\ergsec{\ifmmode {\rm~erg\ s}^{-1} \else
        ~erg s$^{-1}$\fi}
\newcommand\funit{\ifmmode {\rm~erg\ s}^{-1}\;{\rm cm}^{-2} \else
        ~ergs s$^{-1}$ cm$^{-2}$\fi}
\newcommand\phflux{\ifmmode {\rm~photon\ s}^{-1}\;{\rm cm}^{-2}
        \else   ~photon s$^{-1}$ cm$^{-2}$\fi}
\newcommand\efluxA{\ifmmode {\rm~erg\ s}^{-1}\;{\rm cm}^{-2}\;{\rm
        \AA}^{-1} \else ~erg s$^{-1}$ cm$^{-2}$ \AA$^{-1}$\fi}
\newcommand\efluxHz{\ifmmode {\rm~erg\ s}^{-1}\;{\rm cm}^{-2}\;{\rm
        Hz}^{-1} \else ~erg s$^{-1}$ cm$^{-2}$ Hz$^{-1}$\fi}
\newcommand\cc{\ifmmode {\rm~cm}^{-3} \else cm$^{-3}$\fi}
\newcommand\FWHM{\ifmmode {\rm~FWHM} \else ${\rm~FWHM}$\fi}
\newcommand\Lsun{\ifmmode L_{\odot} \else $L_{\odot}$\fi}

\newcommand\hbeta{\ifmmode {\rm H}\beta \else H$\beta$\fi}
\newcommand\Kalpha{\ifmmode {\rm K}\alpha \else K$\alpha$\fi}
\newcommand\nh{\ifmmode N_{\rm H} \else N$_{\rm H}$\fi}

\newcommand{\dMcool}{\ensuremath{\dot{M}_{cool}}}
\newcommand{\Msun}{\ensuremath{\rm M_{\odot}}}
\newcommand\Zsun{\ifmmode Z_{\odot} \else $Z_{\odot}$\fi}

\usepackage{gensymb}

\begin{document}\sloppy

\title{Cool-core, X-ray cavities and cold front revealed in RXCJ0352.9+1941 \\ cluster by Chandra and GMRT observations}


\author{S. S. SONKAMBLE\textsuperscript{1*}, S. K. KADAM\textsuperscript{2}, SURAJIT PAUL\textsuperscript{3,4}, M. B. PANDGE\textsuperscript{5}, P. K. PAWAR\textsuperscript{6} and M. K. PATIL\textsuperscript{2,*} }
\affilOne{\textsuperscript{1}Centre for Space Research, North-West University, Potchefstroom, 2520, North West Province, South Africa\\}
\affilThree{\textsuperscript{2}School of Physical Sciences, Swami Ramanand Teerth Marathwada University, Nanded - 431 606, India\\}
\affilFour{\textsuperscript{3}Manipal Centre for Natural Sciences, Centre of Excellence, Manipal Academy of Higher Education, Manipal, Karnataka 576104, India\\}
\affilFive{\textsuperscript{4}Raman Research Institute, Sadashivnagar, Bengaluru 560080, India\\}
\affilSix{\textsuperscript{5}Dayanand Science College, Barshi Road, Latur - 413 512, India \\}
\affilSeven{\textsuperscript{6}Department of Physics, Deogiri College, Aurangabad - 431005, India\\}


\twocolumn[{

\maketitle

\corres{satish04apr@gmail.com (SSS),  patil@associates.iucaa.in (MKP)}

\msinfo{20 Nov 2023}{8 Feb 2024}

\begin{abstract}
This paper presents a comprehensive analysis of 30 ks  {\it Chandra} and 46.8 ks (13 Hr) 1.4~GHz GMRT radio data on the cool-core cluster RXCJ0352.9+1941 with an objective to investigate AGN activities at its core. This study confirms a pair of X-ray cavities at projected distances of about 10.30~kpc and 20.80~kpc, respectively, on the NW and SE of the X-ray peak. GMRT L band (1.4 GHz) data revealed a bright radio source associated with the core of this cluster hosting multiple jet-like emissions. The spatial association of the X-ray cavities with the inner pair of radio jets confirm their origin due to AGN outbursts.  The 1.4 GHz radio power ${\rm 7.4 \pm 0.8 \times 10^{39} \, erg\, s^{-1}}$ is correlated with the mechanical power stored in the X-ray cavities ($\sim$7.90$\times$ {\rm 10$^{44}$} erg s$^{-1}$), implying that the power injected by radio jets in the ICM is sufficient enough to offset the radiative losses. The X-shaped morphology of diffuse radio emission seems to be comprised of two pairs of orthogonal radio jets, likely formed due to a spin-flip of jets due to the merger of two systems. The X-ray surface brightness analysis of the ICM in its environment revealed two non-uniform, extended spiral-like emission structures on either side of the core, pointing towards the sloshing of gas due to a minor merger and might have resulted in a cold front at $\sim$31 arcsec (62~kpc) with a temperature jump of 1.44 keV.

\end{abstract}


}]


\doinum{12.3456/s78910-011-012-3}
\artcitid{\#\#\#\#}
\volnum{000}
\year{0000}
\pgrange{1--}
\setcounter{page}{1}
\lp{1}

\section{INTRODUCTION}
Combined high-resolution X-ray and radio observational studies of gas-rich cool core clusters have provided us with ample evidence regarding the impact of energy released by the central active galactic nucleus (AGN) on the surrounding intracluster medium (ICM) \citep[see reviews,][]{2012AdAst2012E...6G, 2012NJPh...14e5023M, 2012ARA&A..50..455F}. These evidences have been witnessed in the form of giant cavities, shocks in the X-ray surface brightness, and their coincidence with the radio plasma lobes \citep{2006ApJ...652..216R, 2008ApJ...686..859B, 2010MNRAS.404..180D, 2012A&A...545L...3C, 2013MNRAS.435.1108C}. The close association of the X-ray cavities with the radio lobes in several cool core clusters implies that radio bubbles inflated by the jets from central AGN displaces hot gas in its vicinity and carve cavities or depressions in the X-ray surface brightness \citep{2006MNRAS.366..417F,2009ApJ...705..624D,2006ApJ...644L...9W,2016MNRAS.461.1885V,2017MNRAS.466.2054V,2021MNRAS.504.1644P}.

The superb angular resolution capabilities of {\it Chandra} X-ray telescope have also enabled us to perform a detailed study of previously unseen edges in the hot gas environment, commonly known as shocks and cold fronts. These edges exhibit sharp discontinuities in the surface brightness (SB) and temperature profiles \citep{2002ApJ...567L..27M, 2007PhR...443....1M}. Although shocks mark pressure discontinuities, the pressure profile remains almost continuous across the edge due to a cold front and have been observed in several other cluster environments \citep[e.g.][]{2009ApJ...704.1349O,2010A&A...516A..32G,2010A&A...523A..81D,2013A&A...555A..93E,2013ApJ...770...56G,2016ApJ...826...91A} of both, relaxed and merging. The origin of the cold fronts in relaxed clusters is likely related to gas sloshing induced by off-axis minor mergers \citep{2006ApJ...650..102A,2011MNRAS.413.2057R}, while that in the merging clusters is due to the merger of a gas-rich system easily discernible in the plane of the sky \citep{2001ApJ...562L.153M,2007PhR...443....1M,2011ApJ...741..122O,2005MNRAS.357..801M,2013MNRAS.429.2617O,2017MNRAS.472.2042P,2018MNRAS.476.5591B}. Therefore, cold front study provides us with a crucial probe for the ICM dynamics \citep{2016JPlPh..82c5301Z}. 

This paper presents a combined study of 30 ks \chandra\, X-ray data and 46.8 ks 1.4 GHz GMRT radio data on the cool core cluster RXCJ0352.9+1941 aimed at investigating evidence and energetics of the AGN outbursts operative in the core of this cluster. RXCJ0352.9+1941 is reported to host a centrally concentrated H$\alpha$ emitting gas pointing towards its quiescent undisturbed state \citep{2016MNRAS.460.1758H} with the H$\alpha$ based star formation rate equal to 12.6 \Msun\,/yr \citep{2018ApJ...853..177P}. A radio source of relatively flat spectrum was reported to coincide with the core of this cluster, with its C and X band unresolved radio emission exhibiting a small Giga-Hertz Peaked Source (GPS) like structure and a one-sided tail \citep{2015MNRAS.453.1223H}. A recent X-ray study of this cluster has reported detection of a pair of cavities \citep{2016ApJS..227...31S}, however, their role in the AGN feedback and association with radio emission remained unexplored. This paper is structured as Section \ref{sec2} describes a detailed analysis of the {\it Chandra} and GMRT data on this cluster. The results derived from the morphological and spectral analysis of the X-ray emission are presented in Section \ref{sec3} while Section \ref{sec4} provides a discussion on the results and effectiveness of the AGN feedback. Finally, we summarise our important findings in Section \ref{sec5}

Throughout the paper, we adopt cosmological parameters $H_0$ = 70 km\, s$^{-1}$ Mpc$^{-1}$, $\Omega_m$ = 0.27 and $\Omega_{\Lambda}$ = 0.73;  translating to a scale of 1.98\,kpc\,/arcsec at the redshift $z$ = 0.109 of RXCJ0352.9+1941. All quoted errors for spectral analysis stand for the 90$\%$ confidence level unless otherwise stated. Metallicities were measured relative to the solar metallicity table of \cite{1998SSRv...85..161G}.

\begin{figure*}
\centering
\includegraphics[scale=0.44]{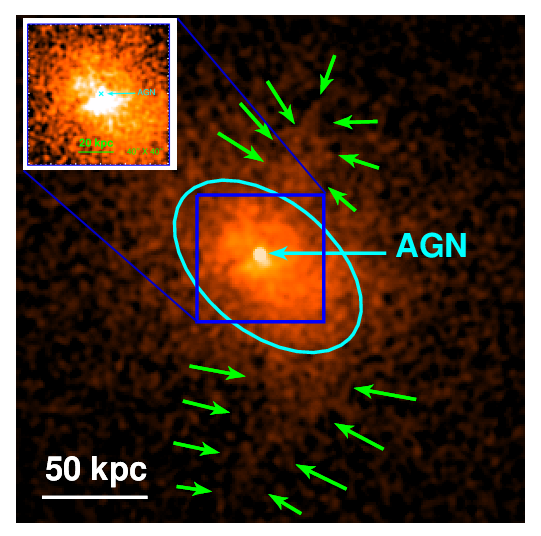}
\includegraphics[scale=0.44]{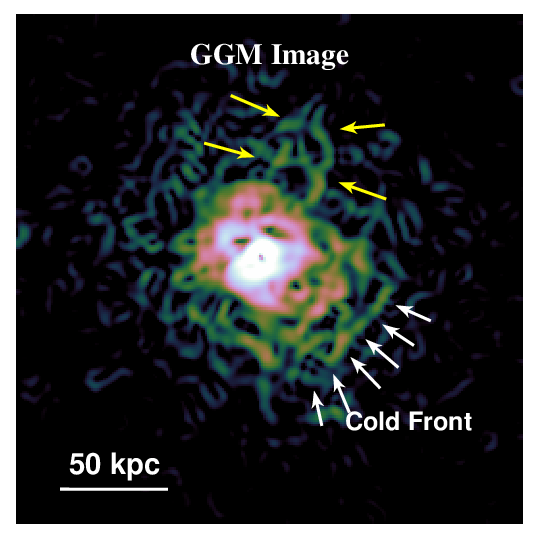}

\caption{{\it Left Panel:} Exposure corrected blank sky background subtracted 0.5 - 3.0 keV 2.0 $\times$ 2.0 arcmin$^2$ \chandra\, image of RXCJ0352.9+1941. For better visibility, the image was convolved with a 2D Gaussian kernel of 1 pixel = 0.492 arcsec. Green arrows on the north and south represent substructures. The white central spot represents the hard X-ray point source (AGN). Magnified view of central region is shown in the inset. The Cyan ellipse presents the extended X-ray emission from the cluster. {\it Right Panel:} GGM filtered image of RXCJ0352.9+1941 on a scale of 3$\sigma$. The edge in the surface brightness is indicated by the white arrows. Yellow arrows in this figure delineate the north substructure in the X-ray emission.}

\label{raw}%
\end{figure*}

\section{Observations and data reduction}
\label{sec2}

\subsection{{\it Chandra} X-ray Data}
RXCJ0352.9+1941 was observed on December 18, 2008, by {\it Chandra} X-ray observatory (OBSID 10466) for an effective exposure of 30 ks ( Table~\ref{obstable}) in VFAINT mode with the object focused on the back-illuminated CCD ACIS-S3. Level-1 event file on this target was retrieved from the \chandra\, Data Archive (CDA) and was reprocessed using \textit{chandra\_repro} routine of {\it Chandra} Interactive Analysis of Observations \citep[CIAO;][]{2006SPIE.6270E..60F} version 4.12 and calibration files CALDB (version 4.9.5). High background flares from the light curve were identified and removed from the event file using the 3$\sigma$ clipping method in \textit{lc\_sigma\_clip} task, which yielded a net exposure time of 27.20 ks. System-supplied background files corresponding to the observations were identified from the \textit{blank-sky} frames using \textit{acis\_bkgrnd\_lookup} task, which were then re-projected to match the observation pointing, roll angle, and were normalised to match the 10 - 12 keV count rates in the science frame \citep{2006ApJ...645...95H}. The CIAO tool \textit{wavdetect} was used to identify point sources within the cluster environment, which were then confirmed by visual inspection, except central, all other point sources were removed from the further analysis. The holes due to removal of the sources were filled in with the background emission employing \textit{dmfilth} tool. For details on X-ray data analysis, readers are referred to \cite{2015Ap&SS.359...61S}.

\begin{figure*}
	\centering
	\includegraphics[scale=0.41]{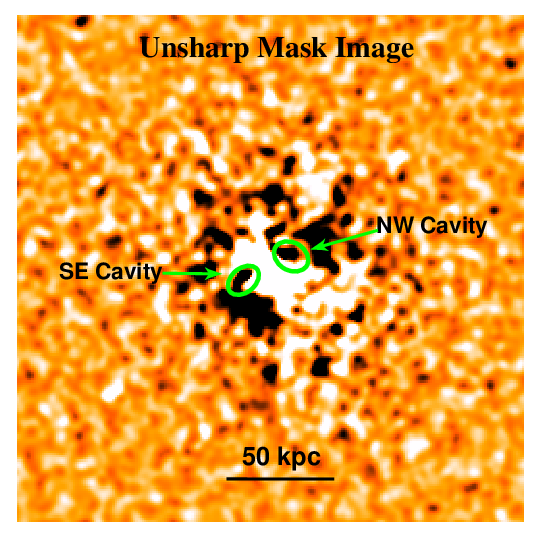}
	\includegraphics[scale=0.41]{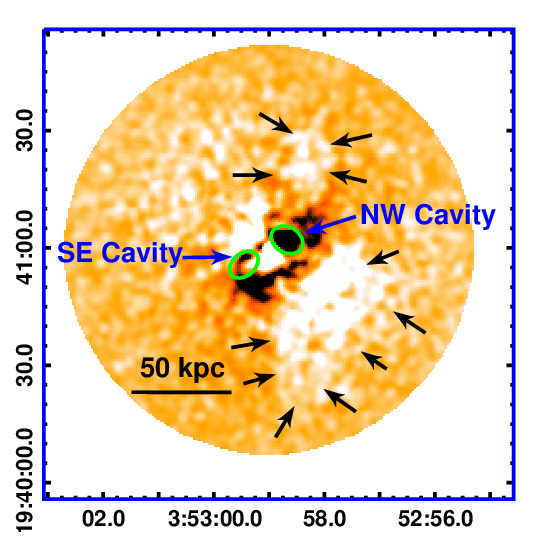}
	\caption{ {\it Left Panel:} 0.5 - 3.0 keV unsharp masked image constructed by subtracting a 10-pixel Gaussian smoothed image from that smoothed with 2 pixels. X-ray deficit cavities are highlighted by green ellipses. Notice the excess emission substructures evident in this image. {\it Right Panel:} 0.5$-$3.0 keV 2D double $\beta$-model subtracted residual map of RXCJ0352.9+1941. Cavities on the SE and NW of the X-ray centre are shown by green ellipses. Black arrows indicate regions of substructures in the form of excess X-ray emission }
	\label{unsharp}%
\end{figure*}

\begin{table}
\caption{{\it Chandra} observation log}
\label{obs}\centering
\begin{center}
\begin{tabular}{lcclcccc}
\hline
ObsId                   & 10466 & \\ 
Observation Date         &  2008 Dec 18 & && \\
Camera                   & ACIS-S  &  &&  \\
Modes (read/data)        & TE / VFAINT &  &&   \\
Exposure (ks)            & 30.00   &         &&  \\
Good Time Interval [GTI]     (ks)             & 27.20   &  && \\
\hline\
\end{tabular}
\end{center}
\label{obstable}
\end{table}

\begin{figure*}
   \centering
   \includegraphics[scale=0.43]{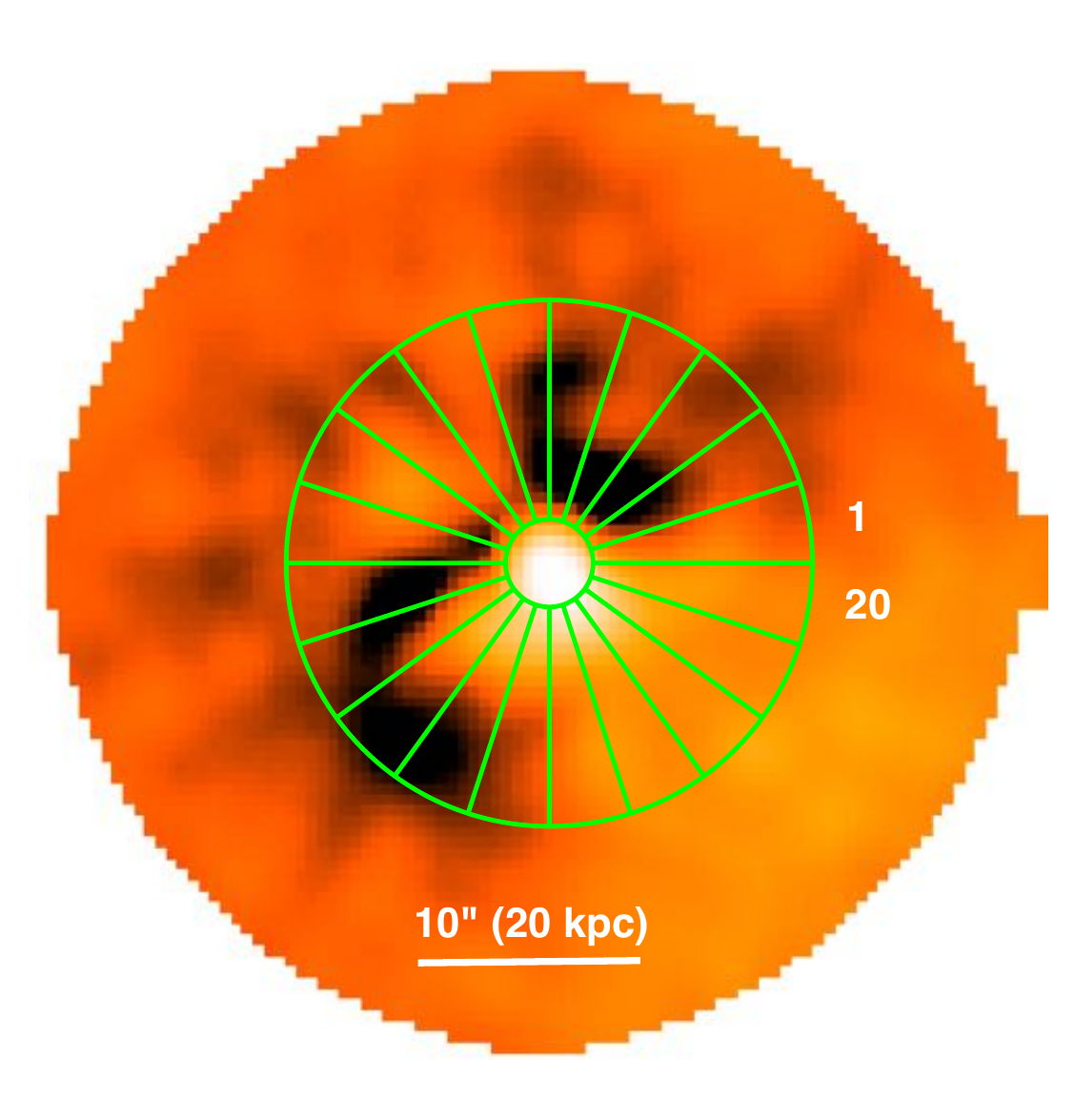}
  \includegraphics[scale=0.55]{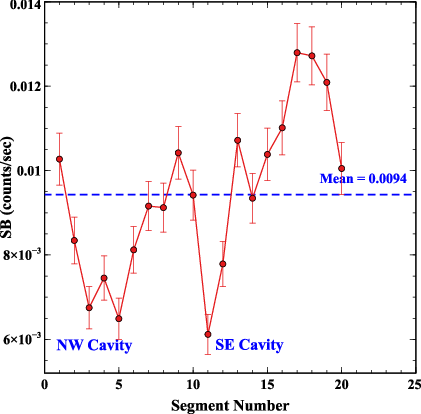}
   \caption{{\it Left Panel:} A 2D double $\beta$-model subtracted residual image overlaid with 20 different segments used to study the variation of counts at the cavity locations. {\it Right Panel:} X-ray count variations are plotted. Notice the dip in counts compared to the mean value shown by a horizontal line.}
       \label{SB_counts}
\end{figure*}

\begin{figure*} 
\centering
\includegraphics[scale=0.57]{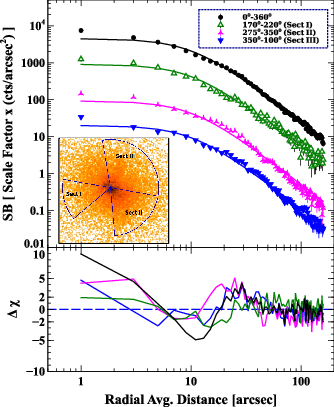}
\includegraphics[scale=0.51]{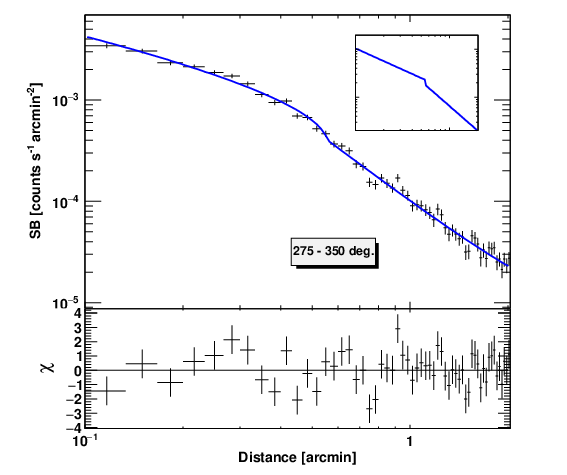}
\caption{{\it Left Panel:} 0.5-3.0 keV azimuthally averaged X-ray surface brightness along with the best-fit 1D $\beta$-model (black filled circles and black solid line). In the same figure, we also plot the surface brightness profiles extracted from the wedge-shaped sectors I, II and III, respectively, covering 170-220\degree, 275-350\degree and 350-100\degree. For better visualisation of these profiles, we add arbitrary offsets in the abscissa. The continuous lines in these profiles show best-fit $\beta$-models. Deviations among the data points relative to the best fit $\beta$-model are shown in the bottom panel. {\it Right Panel:} The surface brightness along sector II  fitted with a broken power-law density model (solid blue line). The corresponding 3D gas density model is shown in the inset, while the lower panel shows residuals of the fit .}
\label{SB_large}
\end{figure*}

\begin{table*}
\caption{Cold front and broken power-law properties}
\centering
\small
\begin{tabular}{ccccccccc}\hline
Region &$\alpha$1& $\alpha$2 & $r_{jump}$ & $n_0$ & Compression &$\chi^{2}$/dof   \\
&&&(arcmin)&($10^{-3}$)& (C) &     \\ \hline
275\degree - 350\degree & $0.95\pm0.02$   &$1.66\pm0.04$  &$0.51\pm0.009$&$1.36\pm0.08$ &$1.37\pm0.05$  &120.26/81   \\  \hline
\end{tabular}
\label{jump}
\end{table*}

\subsection{GMRT Radio Data}
To examine the spatial correspondence between the X-ray deficiencies and the radio emission from the source associated with the cluster, we observed RXCJ0352.9+1941 at 1420 MHz using GMRT \citep{1991CSci...60...95S}. The observations were carried out with 32-MHz bandwidth divided into 512 channels during two observing runs on 22$^{nd}$ and 23$^{rd}$ October 2016 for a total of 13 Hrs (46.8 ks, Project Code 31\_048). The data were recorded in the LL and RR polarisation with an integration time per visibility during both observing runs equal to 16 sec. Standard sources 3C 147 and 0431+206 were also observed during these runs, respectively, for flux and phase calibration. The GMRT data were analysed using the NRAO Astronomical Image Processing Software (AIPS) semi-automated pipeline-based radio-reduction package for flagging and calibration purposes.  After applying the band-pass calibration, the data were averaged using the SPLAT task with care that no bandwidth smearing was happening. The data on the target were then split out using the task SPLIT and were imaged with the task IMAGR and weight parameter robust = 0. Then, amplitude and phase self-calibration were performed after three rounds of phase-only self-calibration. The image {\it faceting} was used to correct the data for the `w' term errors, which were then stitched together using the AIPS task FLATN. Finally, the image was corrected for the primary beam response using the AIPS task PBCOR to achieve the rms noise level of 0.05 mJy beam$^{-1}$.

\section{Results}
\label{sec3}

\subsection {X-ray imaging}\label{Xraymoprph}
Figure~\ref{raw} (left panel) displays an adaptively smoothed, background-subtracted and exposure-corrected 0.5-3.0 keV {\it Chandra} image of RXCJ0352.9+1941. For better representation, the image was convolved with a 2D Gaussian kernel of 1 pixel = 0.492 arcsec. At the centre of the cluster, we find a bright X-ray point source ($\alpha$= 03$^h$52$^m$59\arcsec05, $\delta$=$+$19$\degree$40\arcmin 59\arcsec68), which is surrounded by an extended X-ray emission of diameter $\sim$40 arcsec ($\sim$80\, kpc) shown by the cyan ellipse. In the magnified view of the central region shown in the inset, we see hints of X-ray deficit regions along the SE and NW directions of the X-ray peak. In addition to this, we also find spiral-like emission substructures on the north and the south marked by green arrows. These spiral-like emission substructures suggest that the gas in this system is in a state of sloshing like that evident in several other clusters such as MACS J0416.1-2403 \citep{2015ApJ...812..153O}, MACS J1149.6+2223 \citep{2016ApJ...819..113O}, MACS J0717.5+3745 \citep{2016ApJ...817...98V} and MACS J0553.4-3342 \citep{2017MNRAS.472.2042P}.

With the presence of spiral structure, there is a great possibility of finding edges within the cluster. To reveal this, we constructed a Gaussian Gradient Magnitude (GGM) filtered image of RXCJ0352.9+1941 following \cite{2016MNRAS.460.1898S}. GGM determines the magnitude of surface brightness gradients using Gaussian derivatives. To produce the GGM-filtered image, we used the background-subtracted, exposure-corrected 0.5-3.0 keV \chandra\, image yielding the $\sigma$ = 3 arcsec Gaussian width smoothed image of the magnitude of surface brightness gradient \citep{2016MNRAS.460.1898S}. The resultant GGM image is shown in Figure~\ref{raw} (right panel), which delineates the signatures of disturbed ICM. This image reveals a substructure in the north (yellow arrows) and is consistent with those evident in Figure~\ref{raw} (left panel). The presence of the edge in the surface brightness is highlighted by the white arrows.

To better visualise the substructures in the surface brightness distribution of this cluster we have constructed a 0.5 - 3.0 keV unsharp masked image as discussed in \cite{2013Ap&SS.345..183P}. This was achieved by subtracting a wider 10-pixel Gaussian smoothed image from that smoothed with a 2-pixel Gaussian. The resultant unsharp mask image is shown in the left panel of Figure~\ref{unsharp}, which delineates two X-ray deficit cavities (NW and SE) and excess emission in the north and south directions. 
	
We also obtain a 2D smooth model subtracted residual map (Figure~\ref{unsharp}, right panel) of the cluster emission. In the present case, as the background subtracted, exposure corrected image (Figure~\ref{raw} left panel) and unsharp mask images revealed asymmetries in the X-ray surface brightness which means a single 2D $\beta$-model would not be suitable, therefore, to obtain its smooth model we fitted the X-ray emission with a double 2D $\beta$-model like used by \cite{2015MNRAS.448.2971I}. This was achieved by combining {\tt beta2d} + {\tt beta2d} models within {\it Sherpa} \cite[]{2001SPIE.4477...76F} with Cash statistics. The resultant residual map (Figure~\ref{unsharp}, right panel) showed several interesting features in the central region of RXCJ0352.9+1941, including a pair of X-ray cavities (NW and SE, green ellipses). The size of the cavity is typically determined by visual inspection of X-ray images such as unsharp mask, beta model subtracted residual image, and surface brightness profile \citep{2012ApJ...753...47D}. This method relies on the quality of the X-ray data and is susceptible to systematic errors. Discrepancies in cavity size and shape reported by different authors can vary significantly based on the approach adopted, leading to differences in the measurements \citep[e.g.,][]{2010ApJ...714..758G, 2011ApJ...735...11O, 2012ApJ...753...47D}. Here, we measured cavity sizes from the residual image (8.54 kpc $\times$ 6.56 kpc for NW cavity and 8.22 kpc $\times$ 5.68 kpc for SE cavity) and are comparable to what was reported by \cite{2016ApJS..227...31S}, with an error margin of about 20\%. We also found two additional substructures exhibiting excess emission (highlighted by green arrows) on the north and south of the X-ray centre, as evident in Figure~\ref{raw} (left panel). The substructure on the south appears to be more extended than that on the north.

We also conducted an X-ray count rate variation study to verify the significance of the X-ray surface brightness depressions. We extracted counts from 20 identical segments within annular regions ranging from 2 - 12 arcsec from a background subtracted image in the 0.5 - 3.0 keV range. It should be noted that the $\beta$-model subtracted image in Figure~\ref{SB_counts} (left panel) is only for better visualisation of depressions. The count rate plot in the right panel reveals a decrease in segments 1-7 and 9-13 due to the NW and SE depressions, respectively. These depressions are usually referred to as X-ray cavities. While the residual image shows some additional depression at the position of segments 13 and 14, the count variation does not show signs of an X-ray depression in those sectors. Therefore, this is most likely not a real feature, but an artefact caused by oversubtraction of the model. Similarly, the complex central region of the cluster and its heavier smoothing may also contribute to the generation of such artefacts in the image. Such artefacts have been reported in \cite{2012MNRAS.424.2971O}.

\subsection{X-ray surface brightness profiles}
\label{SB}
To examine the overall morphology of the ICM in this cluster environment, we calculated the azimuthally averaged surface brightness profile of the X-ray emission. This was done by extracting counts from concentric circular annuli centred on the X-ray peak of the background subtracted 0.5-3.0~keV \chandra image of this cluster. The extracted surface brightness was then fitted with the standard $\beta$-model \citep{1976A&A....49..137C} using {\it Sherpa} {\tt beta1d} task \cite[]{2001SPIE.4477...76F}, which resulted in r$_c \sim 10.12\pm0.22$\,arcsec and $\beta \sim 0.54\pm0.002$. The best fit azimuthally averaged (0\degr ~\rm{--} 360\degr) surface brightness profile is shown in Figure~\ref{SB_large} (left panel, dark continuous line), while the data points are shown in dark filled circles. As the central emission could be dramatically different in different systems, the best-fit parameters may not be consistent with those in other groups and clusters. To further examine any discontinuities in the surface brightness we also derive such profiles for the emission extracted from three different wedge-shaped sectorial regions and are shown in the same figure. Here, the profile for the sector I (covering 170 \rm{--} 220\degr) is shown by a green continuous line with data points in green open triangles, sector II (275 ~\rm{--} 350\degr) magenta line and stars, and sector III (350 ~\rm{--} 100\degr) shown with the blue line and data with blue down-triangles. All angles are measured in the counter-clockwise direction. For better visualisation of these profiles, we add arbitrary offsets in the abscissa. The profiles along sectors I and III show small deficits at about 10.5 arcsec (20.80 kpc) and 5.20 arcsec (10.30 kpc), respectively, with significance ranging from 3$\sigma$ to 5$\sigma$ as seen from the bottom of Figure~\ref{SB_large} (left panel). Profiles along sector II and sector III exhibit excess emission between radius of 20\arcsec\, to 40\arcsec\, probably due to the brighter extended emission and the presence of spiral substructure on the north and south. That was observed in Figures~\ref{raw} and~\ref{unsharp} with a significance between 4$\sigma$ to 5$\sigma$. Sector II profile also exhibits an edge (discontinuity) in the surface brightness at about 31 arcsec.

The discontinuity in the surface brightness distribution along sector II was further explored and its geometry was modelled. The centre and shape of the discontinuity were considered as evident in the GGM image shown by white arrows (Figure~\ref{raw}, right panel). We then extracted the surface brightness profile of the X-ray photons along sector II, divided in appropriate bins.  These extractions were then fitted with the deprojected broken power-law density model within PROFFIT V 1.4\footnote{\color{blue} {http://www.isdc.unige.ch/\%7deckert/newsite/Proffit.html}}. This was found to be an effective tool for investigating discontinuity in the surface brightness \citep{2011A&A...526A..79E,2012A&A...541A..57E} and has been used in literature for several systems \citep{2021A&A...650A..44B, 2020MNRAS.491.2605P, 2018MNRAS.479..553S}. This broken power-law model is defined as:

\begin{eqnarray}
    n(r) = \begin{cases} C\,n_{\rm {0}} \left(\frac{r}{r_{\rm jump}}\right)^{-\alpha1}\,, & \mbox{if } r \le r_{\rm jump} \\ n_{\rm {0}} \left(\frac{r}{r_{\rm jump}}\right)^{-\alpha2}\,, & \mbox{if } r > r_{\rm jump} \end{cases} \,,
\end{eqnarray}
where, $n(r)$ represents the electron density at the projected distance $r$, $n_0$ the density normalisation, $C = n_{e_2}/n_{e_1}$ represents the compression factor at the discontinuity, $\alpha$1 and $\alpha$2 the power-law indices on either side of the discontinuity, and $r_{jump}$ the radius corresponding to the putative discontinuity or jump. 

The surface brightness extraction along with the best-fit broken power law are shown in Figure~\ref{SB_large} (right panel). This analysis reveals a break at around 31 arcsec (62 kpc). To ensure that this discontinuity is not an artefact we tried several extraction ranges by varying the angular widths as well as the radial bin sizes as suggested by \cite{2017MNRAS.464.2896C}. Additionally, the task PROFFIT itself ensures proper detection of discontinuity in the selected area. The resultant best-fit analysis, within statistical limits, yielded the power-law indices across the discontinuity as $\alpha$1 = 0.95$\pm$0.02 and  $\alpha$2 = 1.66$\pm$0.04, while the density jump factor as 1.37$\pm$0.05 (Table~\ref{jump}). The discontinuity radius obtained from the fit is $r_{jump}$ = 0.51$\pm$0.009 arcmin. We also tried to check discontinuities in other directions of the cluster emission, however, lower statistics in the present \chandra image failed to find those. 

\begin{figure}[h] 
\centering
\includegraphics[scale=0.65]{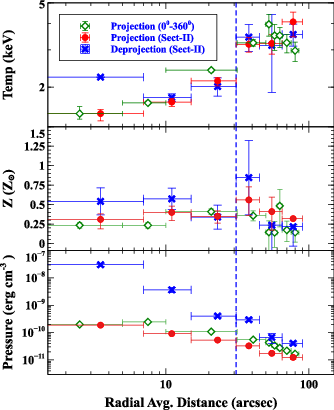}
\caption{Azimuthally averaged projected profiles (green open diamonds) of temperature (upper), metallicity (middle) and pressure (lower) plotted as a function of radial distance. We also plot the projected (red-filled circles) and the deprojected profiles (blue crosses) for the extraction along sector II. The vertical blue dashed line indicates the position of the discontinuity.}
\label{thermo}%
\end{figure}

\begin{table}
\caption{Spectral properties of the nuclear emission}
\begin{centering}
\begin{tabular}{l c c c c c c c}
\hline
Region & Nucleus (2 arcsec) \\
\hline

Model \& Parameters& {\tt tbabs x (apec + pow)}  \\
kT (keV) &   
$1.30^{+0.30}_{-0.27}$   & 
\\
Abundance (\Zsun) & 
0.35 (fixed)&
\\
Norm ($10^{-4}$) & 
$0.20^{+0.20}_{-0.23}$ &
\\
Photon Index ( $\Gamma$ )& 
$0.72^{+0.12}_{-0.15}$&
\\
Norm ($10^{-4}$) & 
$0.18^{+0.05}_{-0.06}$ &
\\
$\chi^{2}_{\nu}$ / d.o.f. &
20.40 / 17 & 

\\
\hline
\end{tabular}
\end{centering}
\label{nuctable}
\end{table}

\subsection{Spectral Analysis of the ICM emission}
\label{section3.3}
\subsubsection{Radial profiles of thermodynamical parameters:}
\label{radilicm}
To investigate the radial dependence of the ICM properties we computed azimuthally averaged projected profiles of thermodynamic parameters such as ICM temperature, metallicity, and pressure. For this we extracted 0.5-8.0 keV spectra from 9 different concentric annuli centred on the X-ray peak of RXCJ0352.9+1941, using task {\tt specextract} within CIAO. The weighted redistribution matrix (RMF) and weighted auxiliary response (ARF) files were derived for each of the extractions. The widths of the annuli were set so that each region had a minimum of $\sim$2000 background-subtracted counts, binned with a minimum of 25 counts per bin. Spectra from each of the annulus were then fitted individually using XSPEC version 12.12.0 \citep{1996ASPC..101...17A} with an absorbed single temperature thermal model ({\tt tbabs x apec}) \citep{2001ApJ...556L..91S,2012ApJ...756..128F}. The Galactic hydrogen column density was fixed at $N_H$ = 1.37$\times$10$^{21}$ cm$^{-2}$ \citep{1990ARA&A..28..215D}, while the redshift was fixed at $z=0.109$. The resultant best-fit parameters, such as temperature and metallicity, were estimated from constrained spectra.

We then compute the electron density $n_e$ (cm$^{-3}$) using {\tt apec} normalisation and the expression provided in \cite{2022ApJ...938...51A}.
\begin{equation}
    n_e  = \biggl[1.2 \ N \times 4.07\times10^{-10}(1+z)^2 \\ \biggl (\frac{D_A}{\rm{Mpc}}\biggl)^2\biggl(\frac{V}{\rm{Mpc^3}}\biggl)^{-1}\biggl]^{1/2},
\label{eq:density}
\end{equation}

where, $D_A$ represent the angular diameter distance, $V$ volume of the spherical shell used for extraction, $N$ the {\tt apec} normalisation in XSPEC, and $z$ the redshift of the object. We assume the ratio of electron to hydrogen density ($n_e/n_H$) equal to 1.2 \citep{1989A&A...215..147B} and compute the pressure and entropy of the gas within each annulus using $p = nkT$ and $S = kT n_e^{-2/3}$ (where $n = 1.92 n_e$ for an ideal gas).

The resultant azimuthally averaged projected profiles of temperature (upper), metallicity (middle) and pressure (lower) as a function of radial distance are shown by green diamonds in Figure~\ref{thermo}. Like several other cool-core clusters, the temperature profile (upper panel) of RXCJ0352.9+1941 takes a minimum value in the core, which then increases in the radially outer part \citep{2009ApJ...705..624D,2009ApJ...693.1142S,2012MNRAS.421..808P,2013Ap&SS.345..183P,2015Ap&SS.359...61S}. This profile also exhibits a discontinuity in temperature at $\sim$31 arcsec (vertical dashed line). A marginal discontinuity was also evident in the pressure profile (lower panel). 

To understand the nature of plasma parameters along the wedge-shaped sector II we also extract spectra from 9 different sectorial regions which are shown by red-filled circles in the same figure. Temperature and pressure profiles of azimuthally averaged and sectorial plots show similar features across the edge. This sectorial plot also shows a jump from $2.27\pm0.1$ keV to $3.24\pm0.2$ keV in the temperature and a marginal jump in the metallicity at $\sim$31 arcsec. A marginal discontinuity in the pressure profile (lower panel) is also evident at this location. 
However, such changes in the pressure profiles at the location of discontinuity have also been reported in some other systems e.g., PKS0745-191 \citep{2014MNRAS.444.1497S}, A2052 \citep{2011ApJ...737...99B}, Virgo \citep{2007ApJ...665.1057F}, and Perseus \citep{2003MNRAS.344L..43F} and may represent a weak shock associated with the AGN feedback. To confirm that the jump in pressure is not due to the projection effect of the brightness distribution, we applied the following deprojection technique. This was done by deriving the three-dimensional structure of the ICM using {\tt deproject}\footnote{\color{blue} {https://deproject.readthedocs.io/en/latest/index.html}} tool and following the ``onion peeling'' method presented by \cite{2003ApJ...585..227B}. This method removes the contamination from the external layers of ICM. Here, we obtained the deprojected temperature profile by extracting X-ray photons from 6 different annuli along the wedge-shaped sector II. The widths of the annuli were adjusted so as to achieve the best S/N. Here, we first fit the X-ray spectrum extracted from the outermost shell with an absorbed {\tt apec} model to obtain the temperature, abundance and normalisation parameters. We then removed the contribution from the outer layer from the successive shells to obtain parameters of the inner shell by adding another {\tt apec} component. This resulted in very few counts, forcing us to increase the bin size to reach the required statistics. We repeated this procedure until we reached the centre of the cluster. The resultant profile of the deprojected values of ICM temperature is shown in the same figure (Figure~\ref{thermo}) and are shown by a blue cross. The temperature and metallicity profiles even in the deprojection also follow the same trend as the projected. Here, a sharp temperature jump of 1.44$\pm$0.53 keV from 2.01$\pm$0.19 keV to 3.45$\pm$0.50 keV is evident at about 31 arcsec. The pressure profile in the deprojection analysis, unlike in the projected case, remains continuous across the edge. Thus, a jump in the temperature with pressure remaining almost constant across the edge confirms its association with a cold front and is discussed separately in Section~\ref{bpl}.

\begin{figure*}
\centering
\includegraphics[scale=0.50]{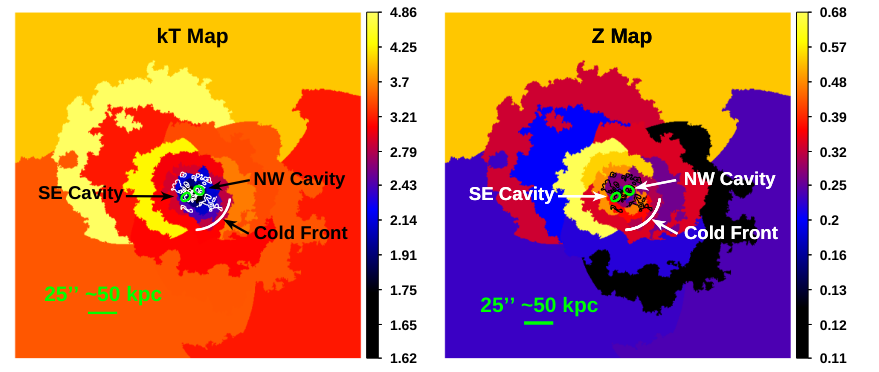}
\caption{2D temperature (left panel) and metallicity (right panel) maps of the ICM obtained from contour binning technique. Overlaid on both these images are the GMRT 1.4 GHz radio contours with levels of $2,8,32,128\times\sigma$. The rms noise is $\sigma=0.05$~mJy~beam$^{-1}$, and the resolution is 2.6 arcsec\,$\times$ 2.0 arcsec\,} 
\label{tempmap}%
\end{figure*}

\subsubsection{2D temperature and metallicity maps:}
\label{text_map}
To understand the 2D spatial variations of temperature and metallicity distribution of ICM in RXCJ0352.9+1941 environment we have computed its 2D maps.  For this we used the contour binning technique (CONTBIN) of \cite{2006MNRAS.371..829S}. This technique identifies the brightest pixels in the point sources removed X-ray image and generates a spatial bin of all such pixels that have the same brightness, which then grows pixel by pixel until it reaches the expected S/N ($\approx$40\,with 1600 net counts). To avoid the formation of elongated bins, we used the geometric constrain factor of C = 2. This yielded a total of 21 bins in 0.5-3.0 keV \chandra\, image. The spectra extracted from each of the bins were then fitted independently adopting $\chi^2$ minimisation. The derived values of temperature and metallicity were then used to plot their 2D maps and are shown in Figure~\ref{tempmap}. The typical errors on the temperature map vary from 5\% in the central region to 13\% in the outskirts, while those in the metallicity map vary from 20\% to 35\%, respectively.\\

The temperature map reveals two arc-shaped regions in the north-east at  50 arcsec (100 kpc) and 114 arcsec (226 kpc) exhibiting the highest value of ICM temperature relative to the ambient gas and are measured to be $4.25\pm0.33$ keV and $4.87\pm0.61$ keV, respectively. It is believed that such arc-shaped patterns in the ICM distribution are the manifestations of the angular momentum of the ICM \citep{2005ApJ...618..227T,2006ApJ...650..102A,2011ApJ...743...16Z,2012A&A...544A.103V}. The temperature map clearly reveals that the coolest gas is segregated in the central region of the cluster implying that the system is a cool core. It is also evident that the cooler, lower entropy gas has elliptical morphology orientated along the northeast to southwest as noticed in Section~\ref{Xraymoprph}. To check the association of the X-ray emitting gas with radio emission, we overlay 1.4 GHz GMRT radio contours on the temperature map, which confirms its spatial association with the X-ray gas envelope. 
The low entropy structures in the central region are likely due to the expanding X-ray cavities in the core region. These might have uplifted the low entropy gas or have triggered its condensation \citep[see,][]{2016ApJ...830...79M,2017ApJ...848...26G,2018ApJ...865...13T}. The metallicity map (Figure~\ref{tempmap} right panel) also reveals arc shaped morphology of high metallicity gas of $0.68\pm0.26$ \Zsun\ at 50 arcsec\, along the eastern direction. This map also reveals asymmetry in the metallicity of gas distributed along the north-east and the south-west directions. These asymmetries in turn suggest that the metallicity mixing due to sloshing is slow and implies that the disturbances caused by the passage of sub-cluster and/or cold fronts are not enough to reach the uniform metallicity mixing. This is in agreement with the findings of \cite{2014A&A...570A.117G} for the Abell 496 cluster.
 
\subsubsection{Nuclear point source emission:}
\label{AGN}
\chandra image of RXCJ0352.9+1941 exhibits a prominent central X-ray source ($\alpha_{\rm J2000.0}$=$03^{\rm h} 52^{\rm m} 59''005 $, $\delta_{\rm J2000.0}$=$+19\degree 40' 59''68$) coinciding with the radio core of the AGN. To understand the emission characteristics of the central source we first find out its spectral hardness  (HR)  = (H - S)/(H + S), where, S and H, respectively, represent the X-ray counts in Soft (0.5 - 2 keV) and Hard (2 - 8 keV) bands extracted from the central 2 arcsec region centred on the X-ray peak \cite{2004ApJ...612L.109W}.  We also estimate the hardness of the surroundings by extracting counts from a circular annulus of width 2 arcsec surrounding the central source. This analysis yielded hardness values of -0.24 $\pm$ 0.04 and -0.40 $\pm$ 0.02, respectively, for the central source and the environment. This clearly exhibits that the central source is much harder than the surroundings thereby confirming the association of AGN with the RXCJ~0352.9+1941. 

For a better understanding of the spectral nature of the central source of this cluster, we also perform spectral fitting of the 0.5-8.0 keV X-ray photons from the same central 2 arcsec region. This could produce a total of 560 background-subtracted counts, which were then imported to the XSPEC and fitted with a combined thermal {\tt apec} and a power-law component. The power law was included to account for the emission from the central hard source as evidenced above. This analysis yielded the best fit temperature value of 1.30$^{+0.30}_{-0.27}$~keV for the ICM metallicity fixed at $Z$ = 0.35$\Zsun$, while the power-law yielded best-fit photon index of $\Gamma$ = 0.72 (Table~\ref{nuctable}). This, in turn, confirms that the central source associated with the cluster is hard enough to deliver (2 - 10 keV) X-ray luminosity of ${\rm L_{X}} = 9.66\times 10^{42} \ \rm{erg \ s^{-1}}$ mainly originating from the non-thermal means.

\subsection{Radio emission features}

\begin{figure}
\includegraphics[width=8.5cm]{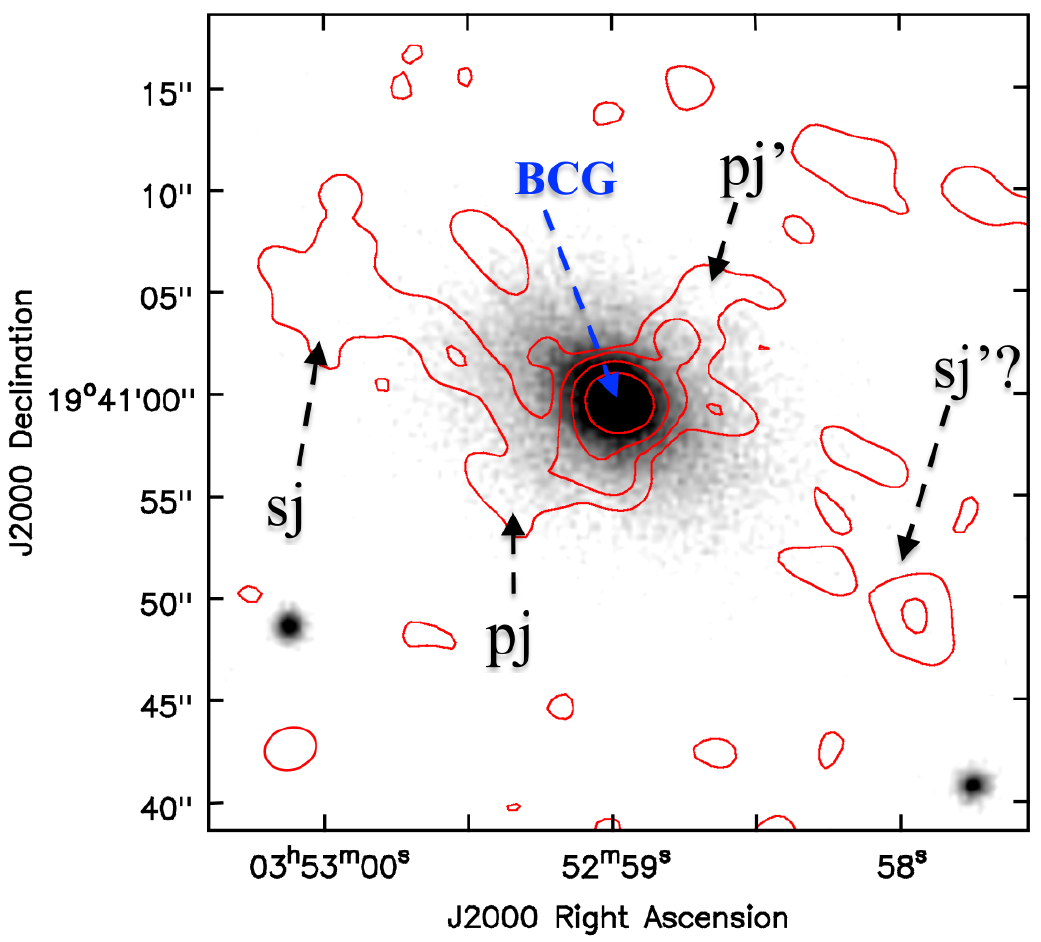}
\caption{GMRT 1.4 GHz radio contours at 2.5,10,40,160~$\times\sigma$ overlaid on the PanSTARSS-1 `r' band image.}\label{fig:radio-opt}
\end{figure}

As reported in \citep{2016MNRAS.461..560G}, RXCJ0352.9+1941 cluster hosts a brightest cluster galaxy (BCG). A radio-loud AGN  associated with the central BCG was detected by the LoFAR telescope at 144 MHz \citep{2020MNRAS.496.2613B}. Additionally, we detect extended radio jet-like diffuse emissions at GMRT 1.4 GHz. To visualize the emission, we overlayed 1.4 GHz GMRT radio contours on the PanSTARSS-1 `r' band image of this cluster (Figure~\ref{fig:radio-opt}). This GMRT image is produced with CASA robust parameter = 0. The final image has an rms of about 50~$\mu$Jy/beam with a beam of $2.6\arcsec\times2.0\arcsec$ and position angle of $71.8^\circ$. The 1.4 GHz contours reveal a strong core with a flux density of 13.7~mJy that co-insides with the central BCG. The radio map also shows the AGN outflow of non-thermal particles in the form of diffuse radio jet-like emission. The longest radio jet (marked as secondary jet or `sj' in Figure~\ref{fig:radio-opt}) appears to be extended up to 20 arcsec\, (40 kpc) towards the northeast direction, along the major axis of the BCG. Although an inner pair of jet-like emissions (marked as primary jet or `pj' and `pj$'$' in Figure~\ref{fig:radio-opt}), along the minor axis of the BCG is also apparent, there is no clear evidence of the counter jet for `sj' along the south-west. Even though a couple of isolated radio sources, possibly due to the counter lobe, are found towards `sj$'$', it cannot be confirmed due to the low fidelity of these sources, as they may also be artefacts. Nevertheless, we compute and report here the total radio emission flux density of the AGN and jets within $3\sigma$ contour as $S_{\rm{1.4}}=20.8\pm2.1$~mJy at 1.4 GHz.

\section {Discussion}
\label{sec4}

\subsection{X-ray cavities as calorimeters}
\label{energetics}

The 2D $\beta$-model subtracted residual map and unsharp mask image revealed a pair of X-ray cavities in the central region of RXCJ0352.9+1941, one on the NW and the SE of its X-ray centre (Figure~\ref{unsharp}). Detection of a pair of X-ray cavities in the environment of this cluster was also reported by \cite{2016ApJS..227...31S}. Assuming that such cavities are the manifestation of AGN outbursts, \cite{2006ApJ...652..216R} have used such cavities as calorimeters and have quantified the mechanical energy injected by the radio jets into the ICM. Like \cite{2016ApJS..227...31S} we assume the ellipsoidal shape of the X-ray cavities carved by the radio jets originating from the central AGN.  The radio lobes of the AGN displaces hot gas carving bubbles or cavities by doing $pV$ work during their outbursts. These cavities then rise buoyantly in the wake of the ICM until they reach pressure equilibrium. At the equilibrium, the buoyant velocity of the cavities exceeds the expansion velocity, get detached from the jets and hence deliver their enthalpy to the ICM. Assuming that the cavities are filled with relativistic plasma, we compute the total enthalpy content of each of the cavities as $E_{cav} = 4pV$ \citep{2004ApJ...607..800B}. Here, $p$ is the pressure of the surrounding ICM and $V$ is the volume of each cavity. Then the cavity power was estimated as 
\begin{equation}
P_{cav} = \frac{E_{cav}}{t_{cav}} = \frac{4pV}{t_{cav}}
\end{equation}

where, $t_{cav}$ represent the age of the cavity and was estimated using the buoyant rise time $t_{buoy} \sim R\sqrt{SC_D/2gV}$ \citep{2006ApJ...652..216R}. Here, $R$ is the projected distance of the cavity from the cluster centre, $g$ the gravitational acceleration ($g = 2\sigma^2 / R$) with stellar velocity dispersion $\sigma$ = 239 km s$^{-1}$ \citep{2018ApJ...853..177P}, $S\, (=\pi R_w^2)$ cross-section area of the cavity, $R_w$ radius of the cavity measured perpendicular to the jet axis and $C_D=0.75$ the drag coefficient \citep{2001ApJ...554..261C}. The plasma pressures $p$ surrounding the X-ray cavities estimated from the projected analysis (Table~\ref{Tab3}) were used to estimate cavity enthalpy. The volume of the cavities was determined using $V = 4 \pi R_l R_{w}^2 /3$ with $R_l$ as the semi-major axis along the radio jet. For the ellipsoidal cavities of sizes given in Table~\ref{Tab3}, we quantify the power content of the NW and SE cavities as 6.01$\times$10$^{44}$ and 1.89$\times$10$^{44}$ erg s$^{-1}$, respectively. Total cavity power of $7.90 \times 10^{44}$ erg s$^{-1}$ corresponds to a net enthalpy content of $\sim6.20\times10^{59}$ erg. Here, uncertainties involved in the estimation of cavity powers depend on the errors in measurement of physical sizes of the cavities, which was done by fitting ellipses to them by visual inspection, and also on the quality of data. As a result, we anticipate greater uncertainties, up to about 20\%, in the enthalpy estimation \citep{2007ARA&A..45..117M, 2010ApJ...714..758G, 2019MNRAS.484.4113K}. \\

\begin{table}
\centering
\caption{Cavity energetics}
\begin{tabular}{@{}lccccr@{}}
\hline 
{\it Cavity parameters}                &    {\it NW-cavity}      & {\it SE-cavity}     \\ 
\hline 
$R_{\rm l}$ (kpc)                     &    6.56            & 5.68              \\

$R_{\rm w}$ (kpc)                     &    8.54            & 8.22              \\

Cavity Vol (cm$^{3}$)                 &   1.06$\times 10^{68}$     & 0.85$\times 10^{68}$         \\

$n_{e}$ (cm$^{-3}$)                    &    0.143                    & 0.132                       \\

$p$ (erg cm$^{-3}$)                    &    2.38$\times 10^{-9}$    & 3.40$\times 10^{-9}$        \\

$t_{buoy}$ (yr)                        &   1.78$\times 10^7$        & 4.73$\times 10^7$            \\

4$pV$ (erg)                            &    3.38$\times 10^{59}$    & 2.82$\times 10^{59}$         \\

$P_{cavity}$ (erg s$^{-1}$)      &    6.01$\times 10^{44}$    & 1.89$\times 10^{44}$     \\
\hline
\end{tabular}
\label{Tab3}
\end{table}

\subsection{Cooling versus heating of the ICM}
\begin{figure}
\includegraphics[scale=0.45]{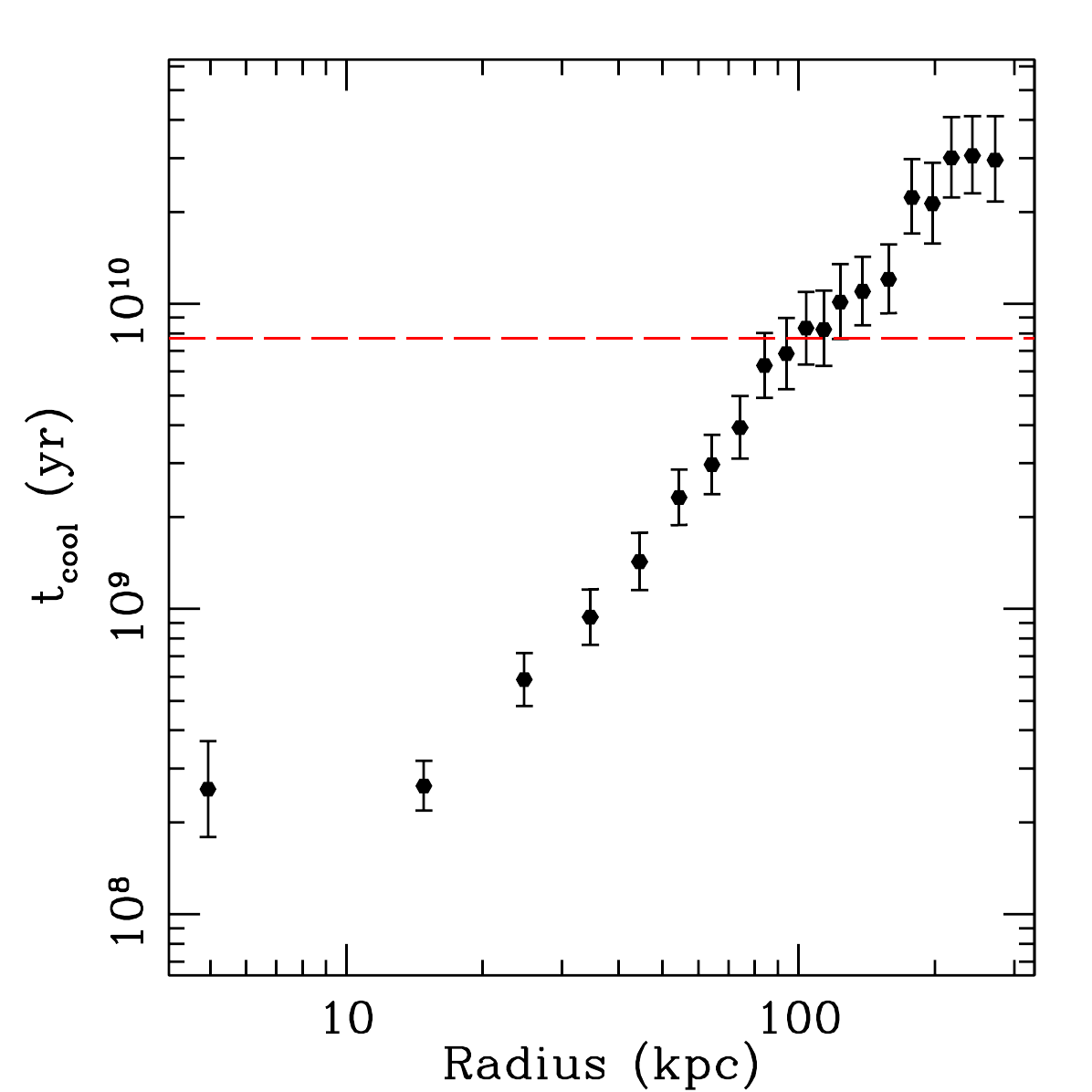}
\caption{Profile of the cooling time of ICM in the core of RXCJ0352.9+1941. The horizontal dashed line corresponds to the cooling time of 7.7 Gyr.} 
\label{cooling_time}%
\end{figure}

In the absence of any central heating, ICM in the core of the cluster must cool radiatively and therefore deposit a large fraction of cool gas at the core. 
From the spectral analysis discussed above, we obtain a profile of ICM cooling time as a function of projected distance using \cite{1988xrec.book.....S}.

\begin{equation}
t_{\rm{cool}} = 0.8 \times 10^{10} \, \mathrm{yr} \left(\frac{n_{e}}{10^{-3} \, \mathrm{cm}^{-3}}\right)^{-1} \left(\frac{T}{10^{7} \, \mathrm{K}}\right)^{1.6}
\end{equation}

where $n_e$ and T, respectively,  represent the electron density and plasma temperature at a projected distance $r$. The resultant cooling time profile is shown in Figure~\ref{cooling_time}, which yields a cooling time of $\sim$ $2.7 \times 10^8$ yr. Assuming classical analogy, we define the ``cooling radius'' (${\rm R_{cool}}$) as the distance where cooling time is less than 7.7 Gyr \citep{2010A&A...513A..37H} (horizontal dashed line) and is equal to 50.33~arcsec\,(100~kpc). We compute the bolometric (0.01 - 100~keV) cooling luminosity within ${\rm R_{cool}}$ as ${\rm L_{cool}} = 1.54^{+0.01}_{-0.01} \times 10^{44} \ \rm{erg \ s^{-1}}$. In a recent study \cite{2023A&A...674A.102W} provided with more physically motivated ways to estimate cooling radii like, cool-core condensation radius (${\rm R_{ccc}}$) and the quenched cooling flow radius (${\rm R_{qcf}}$). These yielded a more tight correlation with the AGN feedback compared to that obtained using the ${\rm R_{cool}}$ value. The estimates for these radii in the present case were found to be ${\rm R_{ccc}}$ = 39.5 kpc and ${\rm R_{qcf}}$ = 58.1 kpc, resulting into the bolometric (0.01 - 100~keV) luminosities equal to ${\rm L_{ccc}} = 6.54^{+0.08}_{-0.10} \times 10^{43} \ \rm{erg \ s^{-1}}$ and ${\rm L_{qcf}} = 9.87^{+0.08}_{-0.13} \times 10^{43} \ \rm{erg \ s^{-1}}$, respectively. The estimates from these radii are found an order of magnitude lower than estimated above using classical cooling radius. \cite{2023A&A...674A.102W} for a sample of cool-core clusters demonstrated that the feeding and feedback processes are linked more tightly when estimated using the ${\rm R_{qcf}}$, even compared to ${\rm R_{ccc}}$.

We then estimate the mass deposition rate adopting classical cooling radius as

\begin{equation}
\dMcool = \frac{2 \mu m_H L_{cool}}{5kT}
\label{eq7}
\end{equation}
where $\mu$ is the molecular weight. This resulted in the mass deposition rate of \dMcool\,= 238$\pm$5.05 \Msun\,/yr, while that estimated using the quenched cooling flow radius (${\rm R_{qcf}}$) was found to be equal to 152$\pm$3.23 \Msun\,/yr. Using the $H_{\alpha}$  flux luminosity reported by \cite{2018ApJ...853..177P} the star formation rate in the core of this cluster to be equal to 12.6 $M_\odot/yr$, an order of magnitude lower than even expected from the quenched cooling flow analogy. This discrepancy between the expected and measured values of the cooling mass confirms that the gas in the core of this cluster is not cooling systematically, but is heated instead to prevent further star formation.

\begin{figure*}
\centering
\includegraphics[scale=0.3]{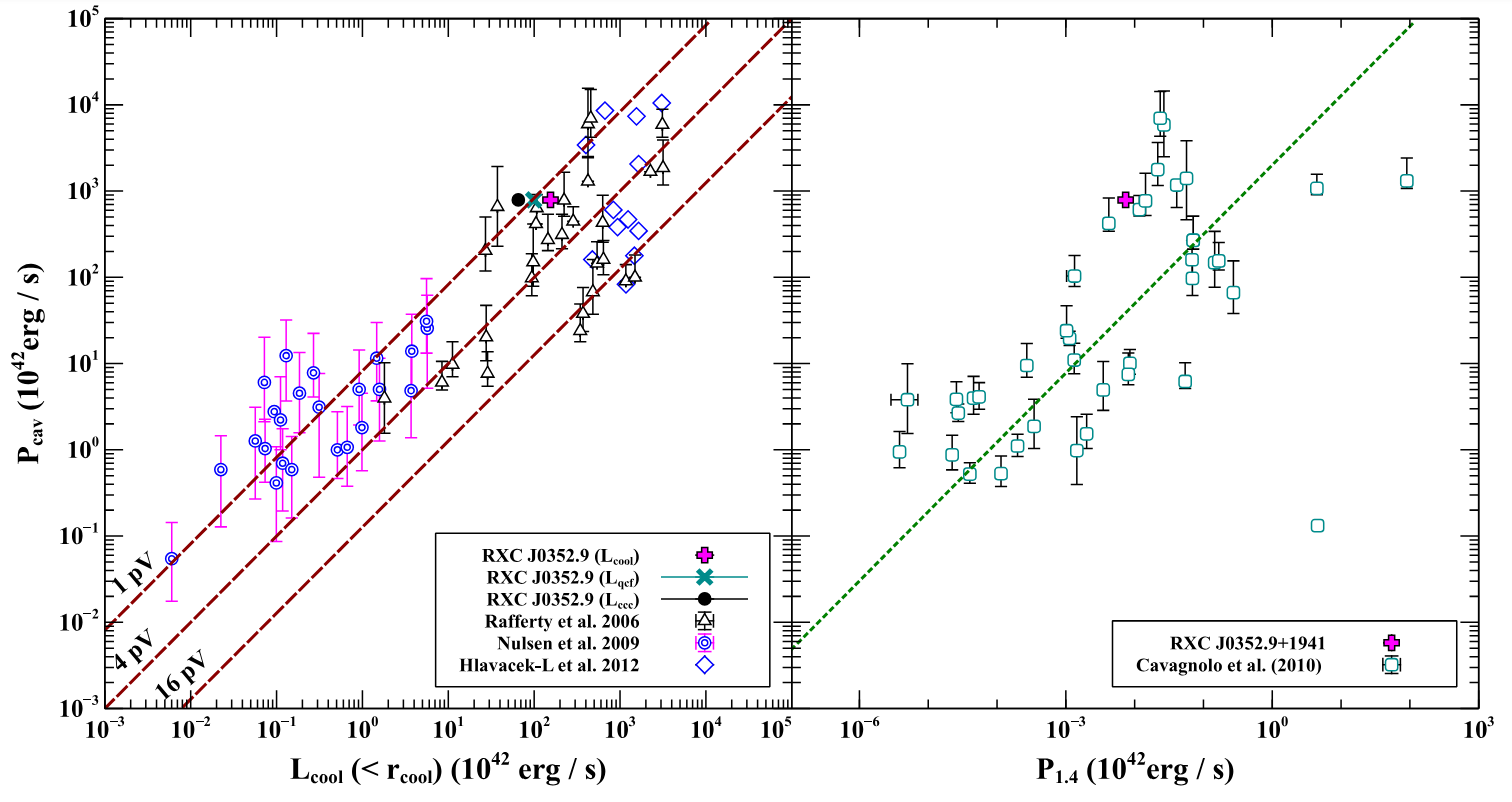}
\caption{{\it Left Panel:} Balance between the mechanical power ($P_{cav}$) versus X-ray cooling luminosity adopted from \protect\cite{2012MNRAS.421.1360H}. We also add data points from the studies by  \protect\cite{2006ApJ...652..216R} and \protect\cite{2009AIPC.1201..198N}. To confirm the balance between the two processes for RXCJ0352.9+1941 we also plot its position estimated using ${\rm R_{cool}}$, ${\rm R_{qcf}}$ and ${\rm R_{ccc}}$ (respectively, magenta plus, green cross and black dot). The diagonal lines from top to bottom, respectively, exhibit the equivalence between the two for enthalpy levels at 1$pV$, 4$pV$ and 16$pV$. {\it Right Panel:} Balance between the cavity power (P$_{cav}$) and 1.4\,GHz radio power, adopted from \protect\cite{2014ApJ...787..134P} for the sample of \protect\cite{2010ApJ...720.1066C}. The green dashed line represents the best-fit relation for the sample of giant ellipticals (gEs). RXCJ0352.9+1941 occupies a position above the best fit of \protect\cite{2010ApJ...720.1066C}.
}
\label{cavrelation}
\end{figure*}

A variety of sources have been considered for the intermittent heating of the ICM. Previous studies on cool-core clusters utilising high angular resolution X-ray data from the \chandra observatory have shown that the enthalpy injected by the radio-jets emanating from the central AGN is capable enough to prevent the ICM cooling and thus, inhibiting the star formation \citep{2006PhR...427....1P, 1994ARA&A..32..277F}. Furthermore, to assess a balance in RXCJ0352.9+1941, we compared the estimates of mechanical power injected by the central AGN using cavities ($P_{cav}$) with the radiative loss of the ICM (${\rm L_{cool}}$). A comparison of the estimates of bolometric cooling luminosity ($L_{cool}$) and total cavity power ($P_{cav}$) confirmed that the enthalpy content of the cavities is
in excess than that required to offset the radiative loss and hence to quench the cooling flow. This was evident from the  position occupied by RXCJ0352.9+1941 in the graph (Figure~\ref{cavrelation} {left panel}) between the cavity power ($P_{cav}$) vs X-ray cooling luminosity ($L_{cool}$) adopted from \cite{2013ApJ...777..163H}. We plot the position of RXCJ0352.9+1941 in this graph by using X-ray cooling luminosities estimated using all the three methods discussed above. The dashed slanted lines in the plot correspond to $P_{cav} = L_{cool}$ for energy inputs of 1$pV$, 4$pV$ and 16$pV$, respectively, from top to bottom. All the three estimates for RXCJ0352.9+1941 hereby confirms that the radio-mode feedback energy delivered by the AGN is enough to compensate the cooling loss, consistent with those seen in other systems studied by \cite{2004ApJ...607..800B}, \cite{2006ApJ...652..216R}, \cite{2009AIPC.1201..198N}, \cite{2006MNRAS.368L..67B}, and at relatively higher redshifts by \cite{2012MNRAS.421.1360H,2015ApJ...805...35H}. The balance between the two evident for the systems belonging to clusters and even for giant ellipticals indicates that the atmospheres of cool core clusters are stabilised by the radio mode feedback.

Combined studies of cool core clusters in X-ray and radio bands have established that the depressions or cavities in the X-ray surface brightness are carved by radio jets emanating from the central AGN \citep[e.g.,][]{2011ApJ...726...86R, 2014MNRAS.442.3192V, 2018ApJ...855...71S}. Further, such cavities are often found to be filled with radio emission of relativistic plasma. In the present case also we found that the extended radio emission at 1.4 GHz mapped using GMRT data covers the X-ray cavities. The measured value of flux density $20.8\pm2.1$ mJy at 1.4\,GHz for RXCJ0352.9+1941 was used to compute its radio power as \cite{2010ApJ...720.1066C}.

\begin{equation}
P_{\nu_{0}}=4 \pi D_{\rm{L}}^{2} S_{\nu_{0}} \nu_{0} (1+z)^{\alpha-1},
\end{equation}

where, $D_{\rm{L}}$, $S_{\nu_{0}}$, and $\alpha$, respectively, represent the luminosity distance, the flux density at frequency $\nu_0$, and the radio spectral index ($S_{\nu} \sim \nu^{-\alpha}$). Assuming $\alpha$ = - 0.8 \citep[typical for radio galaxies; ][]{1992ARA&A..30..575C} 1.4\,GHz radio power was equal to $7.4\pm0.8\times10^{39}$ erg s$^{-1}$. This was then used to check the balance with that of the cavity power P$_{cav}$ as studied by \cite{2010ApJ...720.1066C} (Figure~\ref{cavrelation}~{right panel}). RXCJ0352.9+1941 (magenta plus) occupies a position much above the best-fit relation of \cite{2010ApJ...720.1066C}, implying that the radio source hosted by this cluster is capable enough to deliver sufficient energy and hence to carve the X-ray cavities and to quench the cooling flow. This is in agreement with the results of several other studies \cite{2009ApJ...707.1034B,2010ApJ...714..758G,2011ApJ...735...11O,2019ApJ...870...62P,2016MNRAS.461.1885V,2017MNRAS.466.2054V,2021ApJ...911...66P}. \\

\subsection{Radio jets and cavity association}
\label{radio}
\begin{figure*}
    \includegraphics[width=8.5cm]{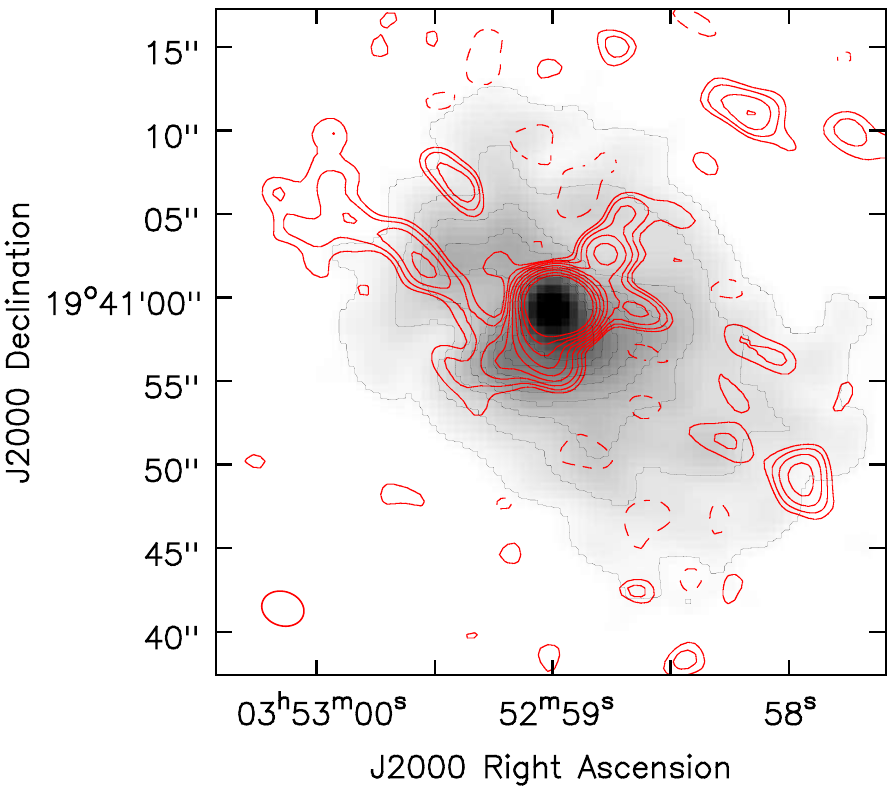}   
\includegraphics[width=8.4cm]{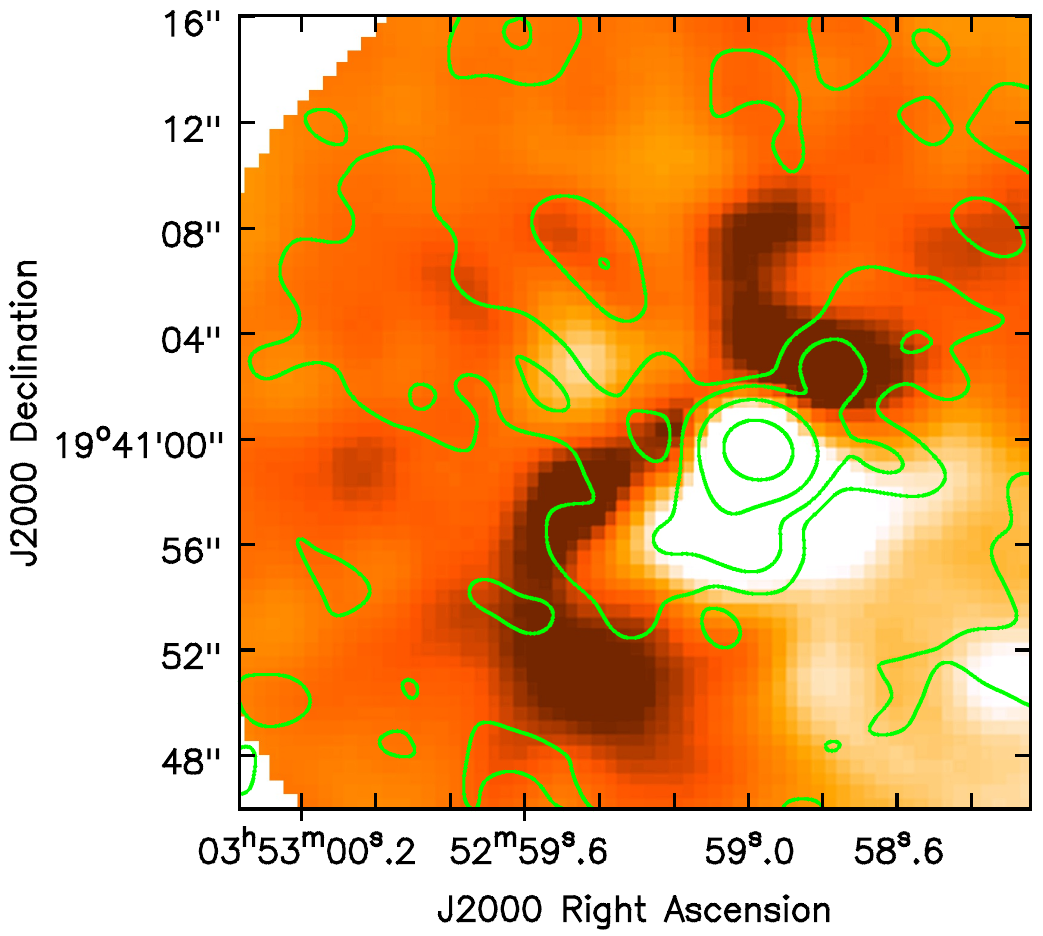}
    \caption{{\it Left panel:} GMRT 1.4 GHz contours (red ) at -3,3~$\times\sigma$ increased by $\sqrt{2}$ times till $48\times\sigma$ overlaid on \chandra\, X-ray emission map (black  contours) at $1-9\times$2 counts. {\it Right panel:} Central 35 arcsec\, region of the $\beta$-model subtracted residual map overlaid with 1.4~GHz GMRT radio contours at $2,8,32,128\times\sigma$. The radio image rms noise is $\sigma=0.05$~mJy~beam$^{-1}$, resolution = 2.6 arcsec\,$\times$ 2.0 arcsec\, and position angle = 71.8$^{\circ}$.}
    \label{fig:radio-xray}
\end{figure*}

A combined study employing data in X-ray and radio bands on a large sample of cooling flow clusters has established a convincing association of the radio source with the core of such clusters \citep{2006MNRAS.373..959D}. RXCJ0352.9+1941 is also reported to host a radio-loud AGN \citep{2016MNRAS.461..560G} that exhibits multiple jets and extended diffuse radio emission like a lobe as we report. Our 1.4 GHz GMRT radio study also confirms the association of a strong radio core with the optical (BCG) and X-ray peak of the central AGN (Figure~\ref{fig:radio-xray} and Figure~\ref{fig:radio-xray} left panel). The diffuse radio emission also reveals the presence of two pairs of jet-like features depicting AGN outflows of non-thermal particles in the form of diffuse jets. The inner pair of jets (`pj' \& `pj$^{\prime}$') coincides with the NW and SE cavities (Figure~\ref{fig:radio-xray}~right~panel), thereby providing an evidence of pushing aside the plasma to carve them. The longest secondary jet  `sj'  is extended up to 20 arcsec\, (40 kpc) in the north-east direction, while `sj$'$' does not provide clear evidence except a couple of isolated sources in the south-west. The diffuse radio-emitting clouds and/or lobes of non-thermal particles were roughly found to occupy regions of low X-ray emission (Figure~\ref{fig:radio-xray},~right~panel).

The radio morphology also exhibits a misalignment between the inner pair of lobes (pj's) and the extended diffuse emission (sj's). Unlike in typical bent-tailed galaxies \citep{2017AJ....154..169S,2020MNRAS.499.5791G,2020AJ....160..161L}, the bending seen in this cluster is not smooth. The sharp bend along with the absence of the counter-lobe makes it difficult to confirm this as a bent-tailed radio galaxy. A strong argument in favour of a bent-jet scenario is the alignment of the base of `sj' with the termination of the `pj' jet and not with the core. This morphology is quite similar to that of the bent-jet radio galaxy in group NGC~1550 \citep{2020MNRAS.496.1471K}. However, the bending angle in RXCJ0352.9+1941 is much larger than in NGC~1550 and unlike in NGC~1550 it does not coincide with the cold fronts evident in this system.

Another viable proposition is that RXCJ0352.9+1941 hosts an X-shaped radio galaxy \citep{2019AJ....157..195L,2019A&A...631A.173B,2020ApJ...889...91G}. The possible mechanisms responsible for the origin of X-shaped radio morphology include precession of jets \citep{1985A&AS...59..511P}, sudden spin flip of jets due to binary black hole merger \citep{2002MNRAS.330..609D}, or diverted back-flow of the ISM/ICM along the minor axis of the host \citep{1984MNRAS.210..929L}. The morphological features evident in the present case such as the absence of hot-spot in the secondary jet (`sj'), significant long length of `sj' relative to primary lobes, and its orientation along the major axis of the BCG all collectively point towards the spin-flip scenario of its formation \citep{2012RAA....12..127G}. However, high-resolution, deep multi-frequency radio data on this source are called for before arriving at any proper conclusion regarding the origin of such an intricate radio morphology.

\subsection{Cold front and sloshing scenario}
\label{bpl}
As discussed in Section~\ref{SB} and \ref{radilicm} the surface brightness analysis revealed a discontinuity in its profile at about 31~arcsec ($\sim$62 kpc) with a density jump of $1.37\pm0.05$. Temperature on the inner (2.01$\pm$0.19 keV) and outer (3.45$\pm$0.50 keV) side of the discontinuity reveals a jump of 1.44$\pm$0.53 keV while pressure maintaining continuity across this discontinuity point towards its association with a cold front like that seen in several other clusters e.g., Toothbrush cluster \citep{2020A&A...634A..64B}, Abell 401, RXC J0528.9-3927 and Abell 1914 \citep{2018MNRAS.476.5591B}, 3C 320 \citep{2019MNRAS.485.1981V}, Abell 2626 \citep{2019MNRAS.484.4113K}, RXJ2014.8-2430 \citep{2014MNRAS.441L..31W}, Abell 496 \citep{2012MNRAS.420.3632R}. 

It is believed that cold fronts in cool core clusters are formed by sloshing of the cluster core, likely triggered by off-axis minor mergers or the passage of small substructures causing the offset in ICM from hydrostatic equilibrium. Such an offset causes the gas in the potential well to oscillate, resulting in the formation of cold fronts around core of the cluster \citep[see,][]{2006ApJ...650..102A,2011MNRAS.413.2057R}. ICM in the environment of RXCJ0352.9+1941 appears highly dynamic, implying that the cold front might have formed due to an off-axis minor merger. The evidence supporting this was provided by the detection of two extended spiral-like features on the north and south part of cluster emission and are consistent with the findings of \cite{2013ApJ...773..114P}.

Sloshing of gas may affect the distribution of relativistic electrons in the cluster due to sloshing-generated turbulence. This may increase radio emission via the synchrotron mechanism \citep{2004ApJ...616..178C, 2019SSRv..215...16V}. Therefore, We speculate that the origin of extended diffuse radio emission surrounding the BCG in RXCJ0352.9+1941 may also be due to sloshing, at least the possibility cannot be fully ruled out. Furthermore, \cite{2005RvMA...18..147F, 2008MNRAS.391.1758K, 2014MNRAS.442..196R} have proposed that an off-axis minor merger may deposit a significant amount of gas in the core of the cluster resulting into an enhanced star formation. In the present case, the reported star formation of 12.6 \Msun\,/yr, higher than that witnessed in other cool core clusters 0.1 to 5 \Msun\,/yr, \citep{2008ApJ...681.1035O,2011ApJ...734...95M}, probably point such a merger.

\section{Conclusions}
\label{sec5}

We conducted a comprehensive analysis of 30 ks  {\it Chandra} data and 46.8 ks (13 Hr)  1.4~GHz GMRT radio data on the cluster RXCJ0352.9+1941 with an objective to investigate AGN activities at its core. We also explore the evidence of AGN feedback and its energy budget in each outburst that it injects into the ICM. Our important findings from the study are summarized below.\\

\begin{itemize}
\item This study confirms a pair of X-ray cavities at projected distances of 10.30 kpc and 20.80~kpc, respectively, on the NW and SE of the X-ray peak and was carried out employing various image processing techniques. Total mechanical power stored in the cavities was estimated to be $\sim$7.90$\times$ {\rm 10$^{44}$} erg s$^{-1}$, while the enthalpy $\sim$6.20$\times$ {\rm 10$^{59}$} erg,  much higher than required for quenching of the cooling flow in its core.\\

\item Spectral analysis of the plasma distributed in this cluster yielded bolometric (0.01 - 100 keV) X-ray luminosity from within the cooling radius ($\sim$100 kpc) equal to ${\rm L_{cool} = 1.54^{+0.01}_{-0.01} \times 10^{44} \ \rm{erg \ s^{-1}}}$ requiring a mass deposition of $238\pm5.05\ \rm{M_\odot \ yr^{-1}}$. This happens to be an order of magnitude higher than the quantum detected in the form of star formation of 12.6 $M_\odot/yr$, implying that the gas in the core is heated instead by an intermittent source like AGN outburst. 

\item Analysis of the GMRT L band (1.4 GHz) data revealed a bright radio source at the core with multiple jet-like emission characteristics. The observed X-shaped morphology of diffuse radio emission is a composite of an orthogonal extended external one-sided jet and an inner pair of jets along the X-ray cavities. The 1.4 GHz radio power $P_{1.4 {\rm GHz}} = {\rm 7.4 \pm 0.8 \times 10^{39} \, erg\, s^{-1}}$ is found to correlate strongly with the mechanical power quantified from the cavity analysis. The clear association of inner jets with the X-ray cavities and the balance between the radio power and enthalpy content of cavities confirms the intermittent heating of the ICM by radio outbursts of central AGN.

\item The hard X-ray emission from the central 2\arcsec\, with a luminosity $\sim$ 9.66$\times$10$^{42}$ erg s$^{-1}$ and a power-law photon index ($\Gamma$) = 0.72$\pm$0.12 suggests its association with AGN.

\item The X-ray surface brightness evidenced two non-uniform, extended spiral emissions structures on either side of the core, pointing towards the sloshing of gas due to a minor merger.  This could result in a surface brightness edge on the southwest due to a cold front at $\sim$31 arcsec (62~kpc) with a temperature jump of 1.44 keV.  \\

\end{itemize}

\section*{Acknowledgements}
SKK gratefully acknowledges the financial support of UGC, New Delhi under the Rajiv Gandhi National Fellowship (RGNF) Program. MBP gratefully acknowledges the support from the following funding schemes:  Department of Science and Technology (DST), New Delhi under the SERB Young Scientist Scheme (sanctioned No: SERB/YSS/2015/000534), Department of Science and Technology (DST), New Delhi under the INSPIRE faculty  Scheme (sanctioned No: DST/INSPIRE/04/ 2015/000108). SP would like to thank the DST-INSPIRE Faculty Scheme (IFA-12/PH-44) for providing the research funding. The authors thank Dhruba Saikia for his careful reading of the manuscript and suggestions. The authors also thank Dr. Biny Sebastian for the helpful discussion on the radio section. The authors thank the staff of GMRT during our observing runs. GMRT is run by the National Centre for Radio Astrophysics of the Tata Institute of Fundamental Research. This publication has made use of the data from the {\it Chandra} Data Archive, NASA/IPAC Extragalactic Database (NED), NASA\rq{}s Astrophysics Database System (ADS), High Energy Astrophysics Science Archive Research Center (HEASARC). We acknowledge the use of Jeremy Sanders' scientific plotting package {\it Veusz}.


\def\nat{NAT}%
\def\aj{AJ}%
\def\actaa{Acta Astron.}
\def\araa{ARA\&A}
\def\apj{ApJ}
\def\apjl{ApJ}
\def\apjs{ApJS}
\def\aap{A\&A}
\def\aapr{A\&A~Rev.}
\def\aaps{A\&AS}
\def\apss{Ap\&SS}
\def\baas{BAAS}
\def\caa{Chinese Astron. Astrophys.}
\def\cjaa{Chinese J. Astron. Astrophys.}
\def\icarus{Icarus}
\def\jcap{J. Cosmology Astropart. Phys.}
\def\jrasc{JRASC}
\def\memras{MmRAS}
\def\mnras{MNRAS}
\def\na{New A}
\def\nar{New A Rev.}
\def\pra{Phys.~Rev.~A}
\def\prb{Phys.~Rev.~B}
\def\prc{Phys.~Rev.~C}
\def\prd{Phys.~Rev.~D}
\def\pre{Phys.~Rev.~E}
\def\prl{Phys.~Rev.~Lett.}
\def\pasa{PASA}
\def\pasp{PASP}
\def\pasj{PASJ}%
\def\sovast{SOVAST}%
\def\ssr{Space Sci. Rev.}
\def\physrep{Phys. Rep.}
\bibliography{mybib,mybib1}


\end{document}